\documentclass{kluwer}    

\usepackage[]{graphicx}
\newdisplay{guess}{Conjecture}

\newcommand{\msol} {M$_{\odot}$}
\def\lesssim{\mathrel{\hbox{\rlap{\hbox{\lower4pt\hbox{$\sim$}}}\hbox{$<$}}}}
\def\gtrsim{\mathrel{\hbox{\rlap{\hbox{\lower4pt\hbox{$\sim$}}}\hbox{$>$}}}}

\begin{document}

\begin{article}
\begin{opening}         
\title{Detecting the progenitors of core collapse supernovae} 
\author{Stephen J. \surname{Smartt}}  
\runningauthor{S.J. Smartt}
\runningtitle{Supernovae Progenitors}
\institute{ Institute of Astronomy,    
 University of Cambridge,
 Madingley Road,         
 Cambridge}
\date{December 1, 2001}

\begin{abstract}
The masses and the evolutionary states of the progenitors of
core-collapse supernovae are not well constrained by direct
observations.  Stellar evolution theory generally predicts that
massive stars with initial masses less than about 30\msol\ should
undergo core-collapse when they are cool M-type supergiants. However
the only two detections of a SN progenitor before explosion are
SN1987A and SN1993J, and neither of these was an M-type
supergiant. Attempting to identify the progenitors of supernovae is 
a difficult task, as precisely predicting the time of explosion of a massive
star is impossible for obvious reasons. There are several different types of 
supernovae which have different spectral and photometric evolution, 
and how exactly these are related to the evolutionary states of the 
progenitor stars is not currently known. I will describe a novel project
which may allow the direct identification of core-collapse supernovae
progenitors on pre-explosion images of resolved, nearby galaxies. This 
project is now possible with the excellent image archives maintained by 
several facilities and will be enhanced by the new initiatives to create 
Virtual Observatories, the earliest of which ({\sc astrovirtel}) is already
producing results. 

\end{abstract}
\keywords{sample, \LaTeX}
\end{opening}           

\section{Which stars go supernovae ?}  

Supernovae of Types\,II and Ib/Ic are thought to occur during
core collapse in massive stars at the end of their lifetimes. 
However the only definite 
detection of a SN progenitor is that of SN1987A in the LMC \cite{white87}, 
which was a blue supergiant (B3I; \opencite{wal89}). The progenitor of 
SN1993J in M81 was possibly identified as a K0\,Ia star \cite{alder94}.
Neither progenitor is consistent 
with the canonical stellar evolution picture, where core-collapse occurs 
while the massive star is an M-supergiant.
We still don't understand the physical mechanisms which underpin the 
different supernovae types, and how these are related to the evolution of 
the progenitor star. There is an understandable 
lack of observational data to constrain the last moments of stellar
evolution. This conference is 
dedicated to understanding the basic building blocks of galaxy evolution, and 
supernova physics is a fundamental input parameter in determining the dynamical
and chemical evolution of galaxies from the first stars in the
Universe to present day gas-rich galaxies. Linking the observed
supernova types to a stars initial mass, metallicity, binarity, environment
and its subsequent evolution is not only important for those of us working on
massive stellar evolution, it will also impact on galaxy evolution as a whole. 

\section{Two recent supernovae with images before explosion }
The very well maintained data archives of the Hubble Space Telescope, 
the Canada-France-Hawaii Telescope, the Isaac Newton Group of Telescopes
and those of ESO contain a vast array of multi-colour
images of late-type galaxies within approximately 20\,Mpc. 
At the spatial resolution of ground-based telescopes, the most luminous 
individual massive stars can be resolved and their photometry accurately 
measured in galaxies within $\sim$8\,Mpc. At the resolution of 
WFPC2 on board HST we can extend these measurement of individual massive
stars to fainter intrinsic luminosities and distances
out to $\sim$20\,Mpc; the Cepheid Key Project is a clear demonstration of 
this \cite{freed2001}. Hence when a bright supernova is discovered in a spiral 
galaxy within $\sim$20\,Mpc there is now a reasonable chance that images 
have been taken of this galaxy either with HST or a ground based facility
- allowing the 
exciting prospect of directly identifying the star which has exploded. 

Towards the end of 1999 two bright supernovae were discovered in the
spirals NGC1637 (7.5\,Mpc) and NGC3184 (8\,Mpc). These events (1999em
and 1999gi) were both Type II-P, with very similar peak magnitudes
($M_{V}\simeq-16$), very similar $\sim$100 day plateaus, and were both
very faint X-ray and radio sources.  By chance there are 
archive images of these galaxies taken several years before explosion by CFHT
and HST. Similar resolution images
taken after explosion have allowed the supernova position to be
precisely determined on the pre-explosion frames. However in both
cases {\em there is no detection of a progenitor star at the SN
position}.  Unfortunately the precursor objects are below the
detection limits (see Fig.\,1 for example of 1999gi).  By measuring
the sensitivity limits of the images, the bolometric luminosity limits
of the progenitors (as a function of stellar effective temperature)
have been determined. These can be plotted on an HR-diagram with
stellar evolutionary tracks, which allows one to estimate the initial
mass of the progenitor (see Fig.\,2 for example). In
\inlinecite{sma2001} and \inlinecite{sma2002} upper mass limits of
9\msol and 12\msol for 1999gi and 1999em respectively were derived,
with uncertainties of $\pm$3\msol.  These SNe are very similar in
their observed characteristics and have rather similar mass limits. In
particular the low values of their X-ray and radio fluxes
(\opencite{pooley} and \opencite{schlegel}) suggest that the
progenitor star had a relatively low mass-loss rate, which is
consistent with the fairly low masses we derive. This is consistent
with the progenitor stars having initial masses between 9-12\msol, and
having exploded as red-supergiants which have undergone normal
mass-loss in the AGB phase.  \inlinecite{sma2002} have speculated that
this type of homogeneous plateau event (which are generally X-ray and
radio faint) could all come from moderate mass 8-12\msol\ progenitors,
and that a Salpeter IMF would suggest that $\sim$50\% of all
core-collapse events should be similar to 1999gi and 1999em if this is
true. To test this hypothesis we require better statistics on the
relative numbers of the SN sub-types {\em and} crucially more 
direct information on progenitors.

\begin{figure*}
\includegraphics[angle=270,width=12cm]{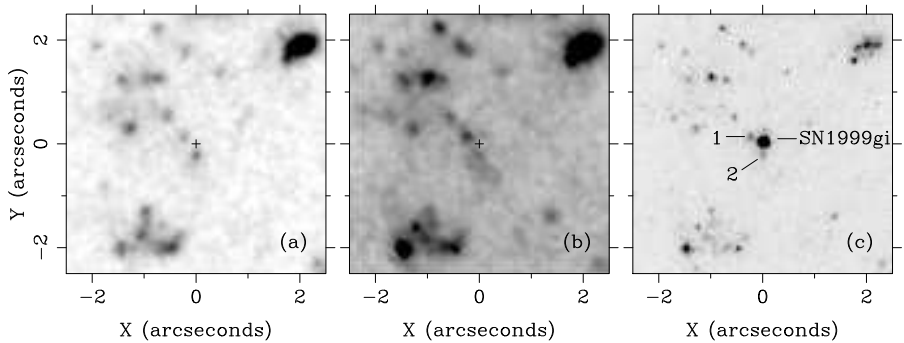}
\caption{The WFPC2 F300W (a) and F606W (b) images before 
the explosion of SN1999gi and the F555W (c) image of the SN  
(from Smartt et al. 2001a). The two luminous
stars OB1-1 and OB1-2 are clearly visible before and after explosion, 
and there is no 
visible object at the position of SN1999gi in either pre-explosion image}
\end{figure*}

\begin{figure*}
\includegraphics[width=10cm]{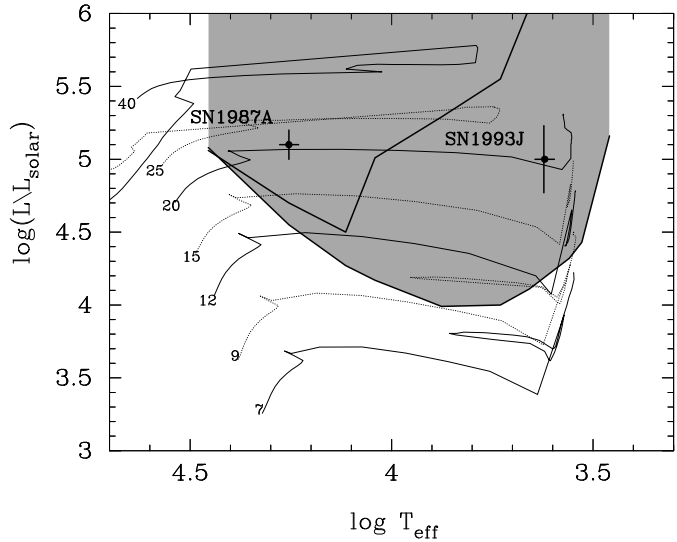}
\caption[]{Mass limits for SN1999gi from Smartt et al. (2001a). 
The Schaller et al. (1992) tracks
for 7$-$40\,M$_{\odot}$ are plotted
with the positions of the progenitor of SN1987A and SN1993J indicated. 
The detection limits of the pre-explosion frames are used to 
set bolometric luminosity limits as a function of stellar effective 
temperature (plotted as the thick solid lines). The WFPC2 pre-explosion
frames should be sensitive to all objects lying in the shaded region
above these lines, with the F300W filter limit being the upper curve.
This indicates an upper limit of 9$^{+3}_{-2}$M$_{\odot}$}
\end{figure*}

\vspace*{-0.2cm}
\section{Future prospects}
The two examples described show how having high-quality archive images of SNe sites
taken {\em prior to explosion} can allow quite stringent limits to be set on the
nature of the progenitor stars. This will be improved in the future if we 
do get a real detection, rather than only a limiting mass. The chances of 
having suitable pre-explosion images available when a nearby core-collapse
supernova is discovered are improving rapidly. By June 2002, a WFPC2 SNAP 
programme will be finished (9042, PI: Smartt) which will enhance the 
number of late-type galaxies with high-quality archive images. Combining these
with data available in the ground-based archives \inlinecite{sma2002} 
have estimated that on average $\sim2.4\pm2$ SNe per year will have 
pre-explosion information and hence a project lasting 3-5 years should 
significantly improve our knowledge. In particular this project has been 
greatly aided by the introduction of the 
{\sc astrovirtel}\footnote{http://www.stecf.org/astrovirtel} 
initiative which is a first step at creating tools for future and more
wide ranging Virtual Observatories. It is an example of exciting science that can 
easily be made possible with the European AVO. This is complemented with 
two HST GO approved proposals in Cycle\,10 and Cycle\,11 
(9042 and 9353) that will provide the accurate astrometry of the supernova
for positioning on the pre-explosion images. Already we have one more candidate 
that has pre-explosion HST data found by {\sc astrovirtel}
(2001du; \opencite{smartt2001}) with follow-up HST ToO images of the SN position
just taken at the end of November 2001. 


\theendnotes

\end{article}
\end{document}